\documentstyle[parkcity]{article}
\frompage{001} \topage{002}                                              %|
\title{\vspace*{1cm}Bulk Hadron Production at AGS and SPS}

\authors{
{Reinhard Stock}\\[2.812mm]
{\normalsize
\hspace*{-5pt} Institute of Nuclear Physics, University of Frankfurt \\
D - 60486 Frankfurt, Germany\\[0.2ex]
}}

\abstract{With new data available from the SPS, at 40 and 80 GeV/A,
I review the systematics of bulk hadron multiplicities, with prime focus
on strangeness production. The classical concept of strangeness
enhancement in central AA collisions is reviewed, in view of the statistical
hadronization model which suggests to understand strangeness enhancement to
arise chiefly in the transition from the canonical to the grand canonical
version of that model. I. e. enhancement results from the fading away of
canonical suppression. The model also captures the striking strangeness
maximum observed in the vicinity of $\sqrt{s} \approx $ 8 GeV. A puzzle
remains in the understanding of apparent grand canonical order at the
lower SPS, and at AGS energies.}
\keyword{}
\PACS{25.75}

\begin{document}
\maketitle

\section{Hadron multiplicity and strangeness enhancement}

The first SPS experiments with $^{32}$S-beams at 200 GeV/A showed
an enhancement of various strange particle multiplicities, chiefly
$K^+, \: \Lambda$ and $\overline{\Lambda}$, relative to pion
multiplicities, in going from peripheral to central S + (S, Ag,
Au) collisions \cite{1}. This observation appeared to be in-line
with the pioneering analysis of Rafelski and M\"uller \cite{2} who
first linked strangeness enhancement to the advent of transition
from the hadronic to a partonic phase. This offered lower
effective $s\overline{s}$ threshold, shorter dynamical relaxation
time toward flavour equilibrium, and an additional, nontrivial
effect of relatively high net baryon number or baryochemical
potential: the light quark Fermi energy levels move up, perhaps
even to the $s$-quark mass at high $\mu_B$, and the Boltzmann penalty
factor for the higher mass $s\overline{s}$ pair creation might be
removed. This latter aspect was mostly ignored in the late 1980's
but receives fresh significance as we become increasingly aware of
the crucial role of $\mu_B$.

It is the purpose of this report to present a sketch of our recent
progress, both in gathering far superior data and in the
understanding of the statistical model that was rudimentarily
anticipated in such early strangeness enhancement speculations.

\begin{figure}[htb]
\begin{center}
%\vspace*{-1.1cm}
                 %\insertplot{fig1.eps}
                   \epsfxsize=11cm
              \epsfbox{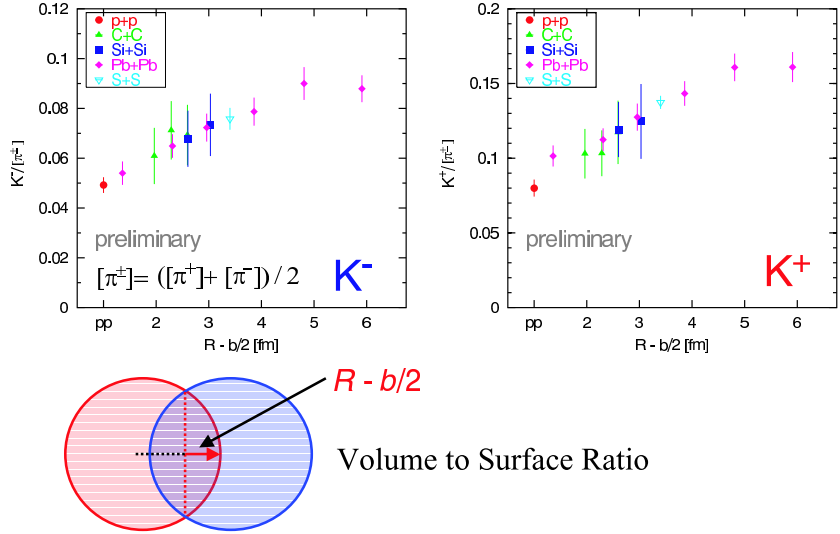}
%\vspace*{-2.1cm}
\caption[]{Negative and positive kaon to pion multiplicity ratios
for Pb+Pb at 158 GeV/A, as a function of collision centrality
given by R-b/2 where R is the nuclear radius. Data for min.
bias pp and for central light nucleus collisions are also given, Ref.3.}
%\label{fig1}
\end{center}
\end{figure}

Fig. 1 shows a modern version of the typical strangeness
enhancement phenomena. Negative and positive kaon to average
charged pion multiplicity ratios obtained by NA49 \cite{3} in
Pb+Pb SPS collisions at 158 GeV/A (corresponding to $\sqrt{s}$=17.3
GeV) are plotted for a sequence of collision centrality
conditions from peripheral to central. At the peripheral end the
minimum bias p+p point \cite{4} matches with the trend. The
centrality scale employed here is unusual but it leads to interesting
conclusions. The raw data bins are ordered in NA49 data by
decreasing projectile spectator energy as recorded in a zero
degree calorimeter. This information is converted to mean
participant nucleon number, or impact parameter b, by a Glauber
calculation. Neither of these scales turn out to be satisfactory
\cite{3} in merging data from central {\it light} nuclei
collisions, C+C, Si+Si and S+S, with the various centralities of
the Pb+Pb collisions. For example, a central S+S collision has $b\approx 2$ fm
and $N_{part} \approx 57$ but on a b scale the $K^+/ \pi$
value is about 40\% lower than the $b=2$ result for the much heavier
PbPb system. Inversely on the $N_{part}$ scale: $N_{part}$=57
corresponds to {\it very} peripheral Pb+Pb and the central S+S result is
about 40\% higher than the Pb+Pb curve. A central collision of a
relatively light nuclear pair thus behaves quite differently from
a very peripheral heavy nuclear collision where only the dilute
Woods-Saxon density tails interact! The scale of Fig. 1 is an
intuitive guess \cite{5} to represent the relative compactness, or
volume-to-surface ratio, by the variable R-$b$/2 where R is the
radius of the colliding nuclear species. It might be connected
with the energy density reached in the primordial collision volume.
We see that the central light nuclear collision data now merge
with the Pb+Pb centrality scale.
Similar NA49 data \cite{6} exist for $\Phi$ and $\overline{K}$
(892) production. The "strangeness enhancement factor" is also
given oftentimes as the production ratio of AA central/(pp min.
bias times 0.5 $N_{part}$). In the case of Fig. 1 it would be roughly two .
Multistrange hyperons \cite{7,8} show factors between 4 and 15.

Bulk strangeness enhancement in central collision is a nuclear feature,
absent in pp collisions. Of course we lack a detailed picture
about "centrality" in pp collisions but we could still employ e.
g. the total charged particle multiplicity to select more or less
"violent" collisions. Fig. 2 shows the $K^+/\pi$ ratio of pp at
158 GeV versus charged particle multiplicity, to be essentially
flat \cite{9}. Similar findings are made up to Tevatron $\sqrt{s}=1.8$
TeV $p\overline {p}$ collisions \cite{10}: the $K^+/\pi$ ratio is
50\% higher here but also independent of $N_{ch}$.

\begin{figure}[htb]
\begin{center}
%\vspace*{-1.1cm}
                 \insertplot{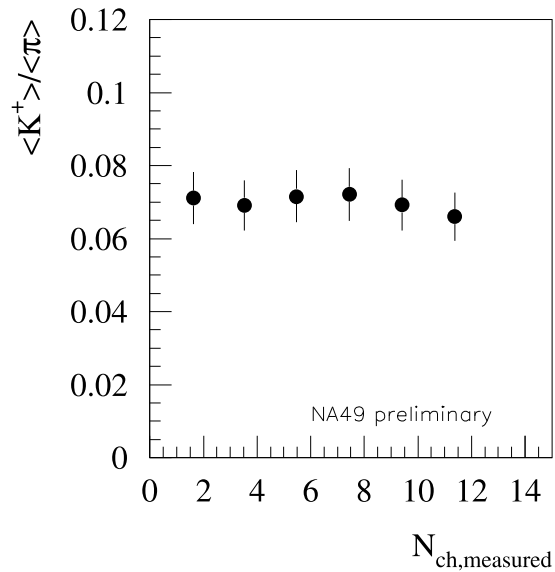}
\vspace*{-0.3cm}
\caption[]{The multiplicity ratio of positive kaons to pions in pp
collisions at 158 GeV, as a function of charged particle multiplicity, Ref.9.}
%\label{fig1}
\end{center}
\end{figure}

A picture emerges in which strangeness enhancement, or more
generally speaking the yield order in the overall bulk hadron
population is connected with "sequentiality" of interactions at
the microscopic level, i.e. with the number of successive
collisions if one may employ a naive Glauber picture: with the
size and density of the primordial interaction zone. Unfortunately
this formal statement does not give us much deeper insight because
if we knew how to describe a second, third etc. collision of a
hadron, within fm/c space-time distance we would have probably
resolved the key issue: does it dissolve into a parton cascade
from which the final hadrons are reconstituted? Proton-nucleus
collisions {\it must} hold a key to this question but nobody has
 succeeded in isolating the second, third, n'th successive
collision of the projectile, as of yet \cite{11,12}.
\begin{figure}[ht]
\vspace*{-1.1cm}
                 %\insertplot{fig3.eps}
                 \begin{center}
                 \epsfxsize=8cm
                \epsfbox{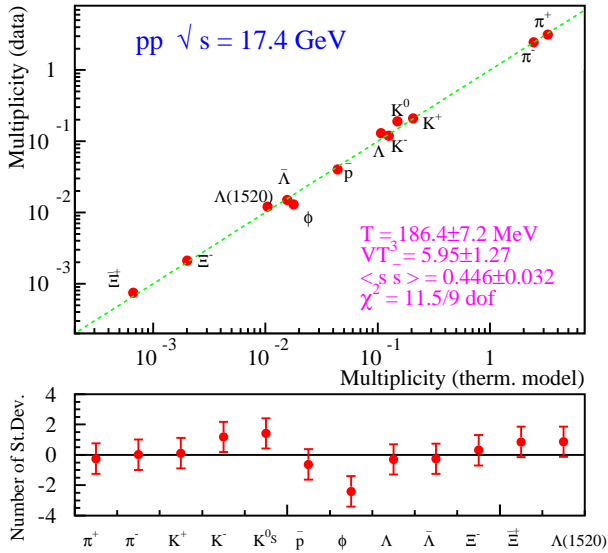}
                \end{center}
\vspace*{-2.1cm}
\caption[]{NA49 data for hadron multiplicities in pp collisions at
158 GeV confronted with the canonical model of Becattini, Ref.15.}
%\label{fig1}
\end{figure}

At the moment we thus forgo pA as an intermediate step although it
certainly also features changes in the hadronic production ratios
\cite{13} and base the analysis on comparing pp to AA. Fig. 3
shows the hadronic multiplicities,  from pion to cascade hyperon,
obtained by NA49 for min. bias pp at $\sqrt{s}$=17.3 GeV
\cite{14}. The data are confronted with the Hagedorn statistical
model in its canonical Gibbs ensemble form as employed by F.
Becattini \cite{15}, leading to very good agreement (as it is well
known also for other elementary  collisions and energies \cite{16}). The three
parameters are T=186$\pm$7 MeV, a reaction volume of 6 fm$^3$, and
a total of about 0.5 $s \overline{s}$ pairs. The apparent validity of a
statistical weight-dominated picture of phase-space filling is
not well understood already since Hagedorn's time. It is clear,
however, that the apparent canonical "hadrochemical equilibrium"
pattern can {\it not} result from "rescattering" of produced
hadrons: there is none. In Hagedorns view \cite{17} a creation
"from above" must hold the key to the apparent maximum entropy
state, i.e. the QCD process of hadronization \cite{18}. This
pattern and T-value are a fingerprint of QCD hadronization - do AA
collision data at high $\sqrt{s}$ also confirm this picture (they must,
of course, also result from a hadronization process)?

\section{AA collisions in the Grand Canonical Model}

Fig. 4 shows the grand canonical fit by Becattini to the NA49 data
from central Pb+Pb at 158 GeV/A \cite{18}. The temperature is 160
$\pm$ 5 MeV and $\mu_B$ =240 MeV; besides, this model employs the much
discussed strangeness undersaturation factor $\gamma_s$=0.8.
\begin{figure}[htb]
\begin{center}
%\vspace*{-1.1cm}
                 \insertplot{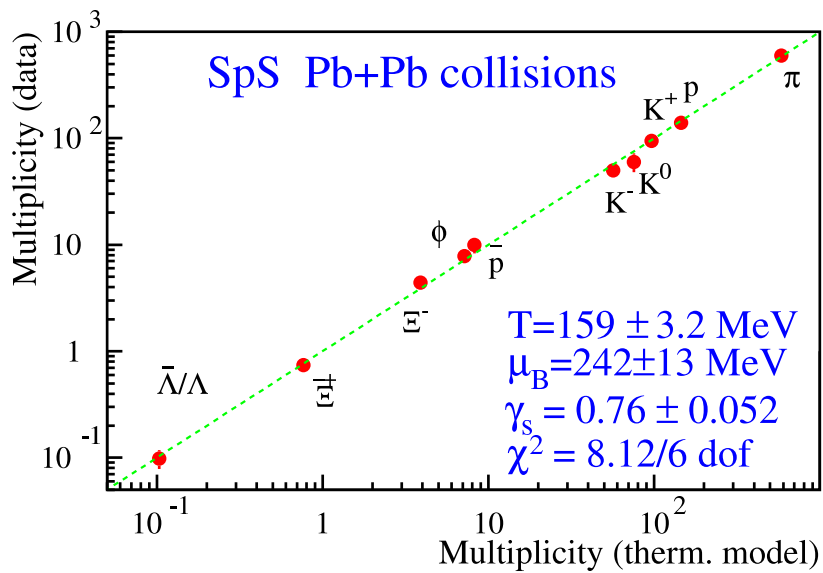}
%\vspace*{-2.1cm}
\caption[]{Hadron multiplicities for central PbPb collisions
at 158 GeV/A from NA49 confronted with the grand canonical statistical
model, Ref.18.}
%\label{fig1}
\end{center}
\end{figure}

Leaving the second order concern about $\gamma_s$ to the
theoretical community I note here that Braun-Munzinger et al.
\cite{19} fit a set of data at the same top SPS energy without
introducing a $\gamma_s$; they report T=170 $\pm$5 MeV, at
$\mu_B$=270 MeV, close enough. There are also studies
 of the new RHIC STAR data \cite{20} at $\sqrt{s}$=130 GeV by
this model \cite{21} and by Kaneta and Xu \cite{22}, averaging at
175 $\pm$ 5 MeV and $\mu_B$=48 MeV. And the new, still preliminary
data of NA49 \cite{23} at 80 and 40 GeV/A have resulted in
Becattini fit values of T=155 MeV, $\mu_B$=270 MeV and T=150 MeV,
$\mu_B$=395 MeV, respectively. I will return shortly to a further
discussion of the grand canonical approach but wish to, first of
all, show an overall impression from these analyses which are
confronted in Fig. 5 with the sensational new lattice QCD
calculations at finite $\mu_B$ by Fodor and Katz \cite{24}.

\begin{figure}[htb]
\begin{center}
%\vspace*{-1.1cm}
                 \insertplot{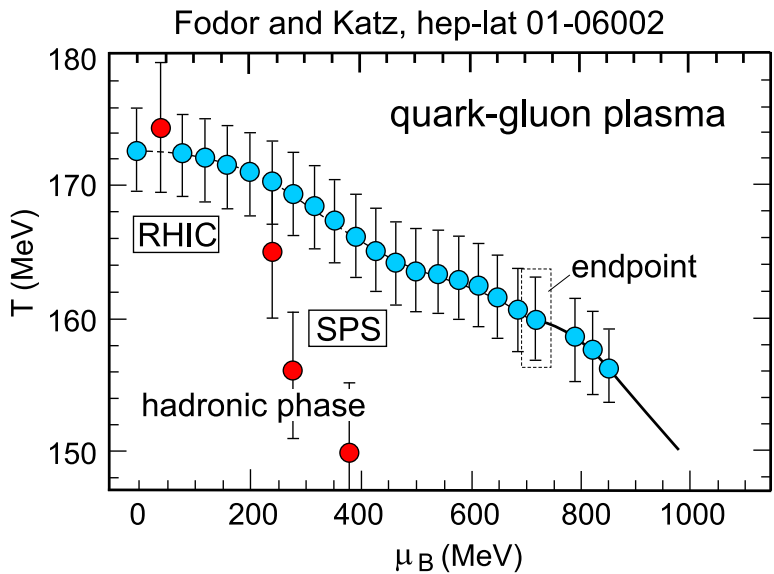}
%\vspace*{-2.1cm}
\caption[]{The lattice QCD phase boundary in the plane of $T$ vs.
$\mu_B$, Ref.24. The hadronization points captured by grand canonical
analysis for SPS and RHIC energies are also shown.}
%\label{fig1}
\end{center}
\end{figure}

The latter predict the T, $\mu$ dependence of the QCD phase
transformation which in this model consists of a crossover for all
$\mu>$ 650 MeV, i.e. in the SPS-RHIC domain. Note that physics
observables can change rapidly in a crossover, too: the familiar
steep rise of e.g. lattice $\epsilon/T^4$  at $T_c$ does not, by
itself, reveal the order of the phase transformation \cite{25}.
Anyhow: the hadronization points from grand canonical ensemble
analysis merge with the phase transformation site of lattice QCD
at top SPS and RHIC energy.
Quite a sensational result, but also a plausible one \cite{26} if
we recall that Ellis and Geiger did already point out in 1996 that
hadronic phase space weight dominance appears to result from the
colour-flavour-spin-momentum "coalescence" that occurs at
hadronization \cite{27}. Unfortunately a rigorous QCD treatment of
the parton to hadron transition is still missing.

At this point the following objection is always raised: if the
same basical model describes hadronic yield ratios in $pp$, $e^+e^-$
and in central AA collisions, Figs. 3 and 4, what is special about
AA, as you will not tell us now that a QGP is also formed in pp?!
Answer: on the one hand both collision systems reveal the QCD
hadronization process which features, furthermore, the Hagedorn
limiting hadronic temperature $T_H$. At top SPS and at RHIC energy
$T$ (hadronic ensemble) $\approx T_H \approx T_c$ (QCD), {\it
this} is the common feature; it should not be a chance
coincidence. On the other hand hadronization appears to occur
under dramatically different conditions in AA collisions,
as captured in the
transition from a canonical to a grand canonical description.
Inspection of Fig. 3 and 4 shows that the hadronic population
ratios are quite different: the falloff from pions to cascade
hyperons in the former case is about four orders of magnitude
whereas it reduces to three in the grand canonical situation:
strangeness enhancement! In the canonical case the small reaction
fireball volume is strongly constrained by local conservation of
baryon number, strangeness neutrality and isospin whereas these
constraints fade away in the GCE which represents a situation in
which, remarkably, these conservation laws act only {\it on the
average}, over a rather large volume, as captured by a collective
chemical potential $\mu=\mu_B + \mu_s + \mu_I$. On top of this, $\mu_s$
is eliminated by {\it overall} strangeness neutrality, and $\mu_I$
by overall charge conservation. This leaves one global quantity $\mu_B$
essentially in charge of all the conservation task. Note that the
statistical model does {\it predict} nothing, it {\it merely
captures} this most remarkable feature of the data. Its observed
success implies some kind of long range collective behaviour in
the hadronizing source, the origin of which is yet unknown, but
must be specific to central AA collisions. Strangeness enhancement
is the fading away of canonical constraints, in the terminology of
the statistical model \cite{28}.
\begin{figure}[htb]
\begin{center}
%\vspace*{-1.1cm}
                 \insertplot{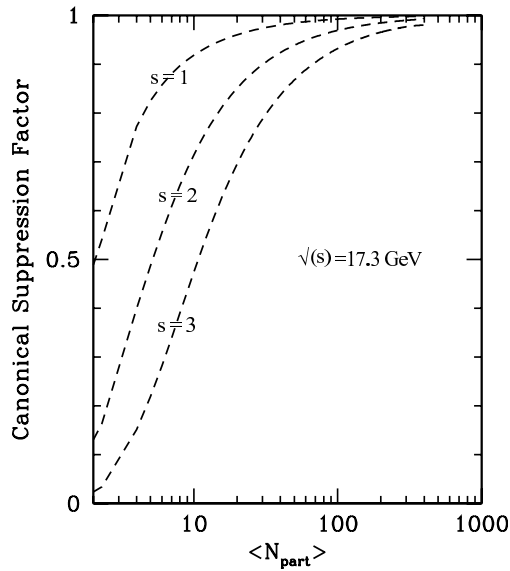}
%\vspace*{-2.1cm}
\caption[]{The canonical to grand canonical transition as reflected
in the canonical suppression factor which is the inverse of strangeness
enhancement, shown for strange hadron species with s=1,2,3 at top SPS
energy, Ref.29.}
%\label{fig1}
\end{center}
\end{figure}

This aspect has been recently studied in all detail by Cleymans,
Redlich, Tounsi and collaborators \cite{28,29}. Fig. 6 illustrates
their results concerning the transition from canonical to grand
canonical behaviour with increasing number of participants, i.e.
overall "source" size. It is intuitively clear that it should
occur, first, in singly strange hadrons, the increase occuring
with offset (but having a larger specific effect on the yields per
participant) in S=2,3 hyperons.

A further, appropriate critical question: how can we understand
the other aspect of Fig. 5, i.e. the steep falloff from the QCD
transition domain occuring at the lower SPS energies? We even have
a further GCE analysis, at top AGS energy, by Braun-Muzinger et
al. \cite{21}, for central Si+Au collisions at 14.6 GeV/A, shown
in Fig. 7.
\begin{figure}[htb]
%\vspace*{-1.1cm}
              %   \insertplot{fig7.eps}
              \begin{center}
              \epsfxsize=10.5cm
              \epsfbox{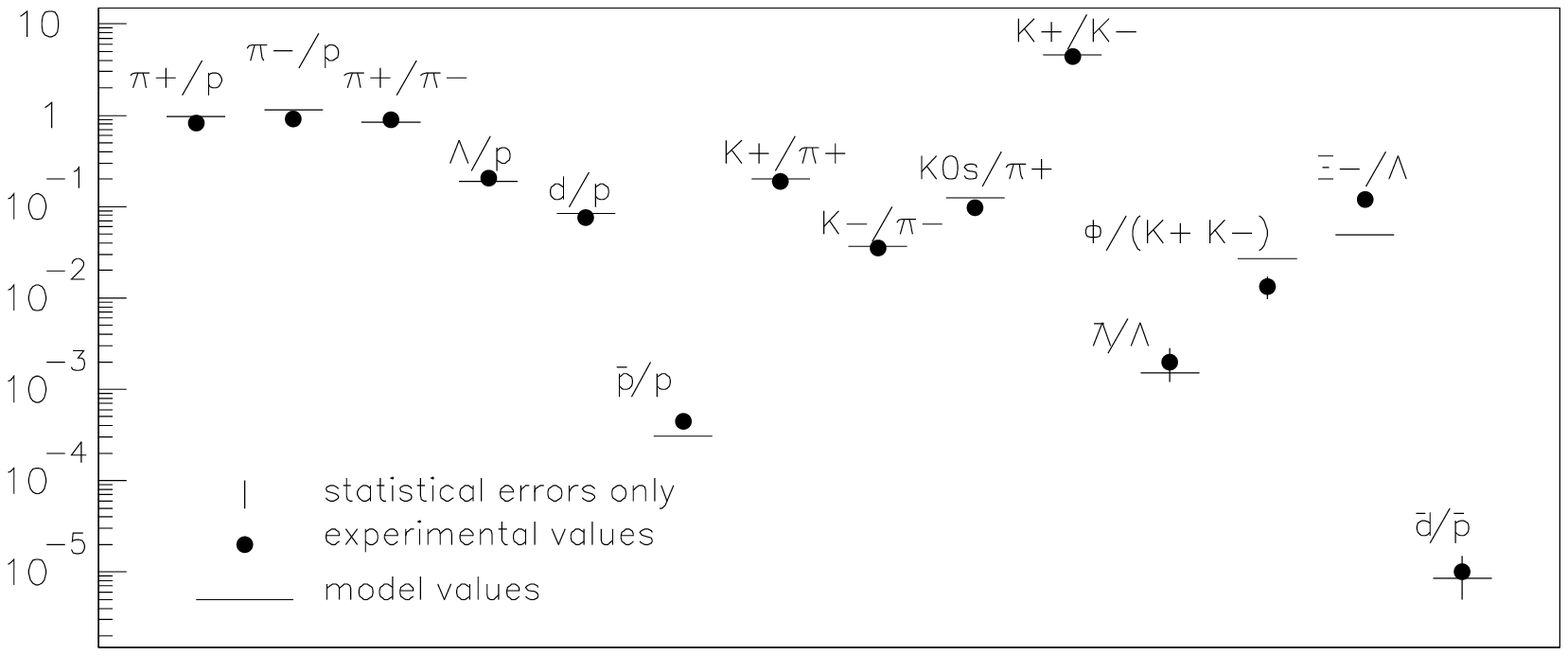}
\end{center}

%\vspace*{-2.1cm}
\caption[]{Hadron yield ratios at top AGS energy, in central
Si+Au collisions at 14.8 GeV/A as fitted with the grand canonical
statistical model, Ref.21.}
%\label{fig1}
\end{figure}

The result is $T$=125 MeV, $\mu_B$=540 MeV, far below
the $T$ scale of Fig. 4. The picture of a direct parton to hadron
transition is intuitively inapplicable at these lower energies.
Still the overall dynamical trajectory that ends in hadronic
chemical (abundance) freezeout should arrive there "from above" as
hydrodynamical models \cite{30} show. How can the expanding
hadronic system maintain conditions near equilibrium, or acquire
them? We do not know. Clearly a primitive hadronic point-like gas
would not accomplish this, due to relaxation times far exceeding
the expansion time scale (volume doubling occuring every about
4fm/c). However, the hadronic system is very dense along its
trajectory, and it is thus a quantum mechanical coherent state
that decays to the finally observed hadron ensemble. Unlike a
quasi-classical, albeit dense "gas" it might
thus ignore the classical concept of a relaxation time. Recall the
nucleus, also still a dense system: we do not invoke relaxation
time in a transition within such  a quantal medium, such as
$\beta$-decay. And yet "Fermis Golden Rule" asserts that the
transition strength depends "only" on the squared matrix element
times final state phase space volume weight plus global
conservation laws. And we know that the phase space factor
oftentimes far overrides the matrix element, in the net decay strength.
Vague hints, at present! High density hadronic matter behaviour is
essentially unknown: an old and new research paradigma. At top SPS
and at RHIC energy, in turn, the increasingly "explosive" nature of
hadronic and partonic expansion may almost instantaneously dilute
the hadronizing source toward chemical freezeout, as indicated
by $T(GCE)\approx T$ (Hagedorn)
$\approx T_c$ (QCD).

\section{Energy dependence of strangeness yields}

From combination of AGS, SPS and RHIC hadron multiplicities we can
construct the energy dependence of various strange particle
yields relative to the pion yield which carries the main fraction
of light quark production. As an example Fig. 8 shows the total $\Lambda$
and cascade hyperon yield ratio \cite{31} relative to $\pi^+$, as
a function of $\sqrt{s}$. A distinct maximum is visible in the $\Lambda/\pi^+$
yield ratio. Similar maxima occur e. g. in the $K^+/\pi$ ratio
\cite{23}, at $\sqrt{s} \approx 6-8$ GeV.
\begin{figure}[htb]
%\vspace*{-1.1cm}
\begin{center}

                 \insertplot{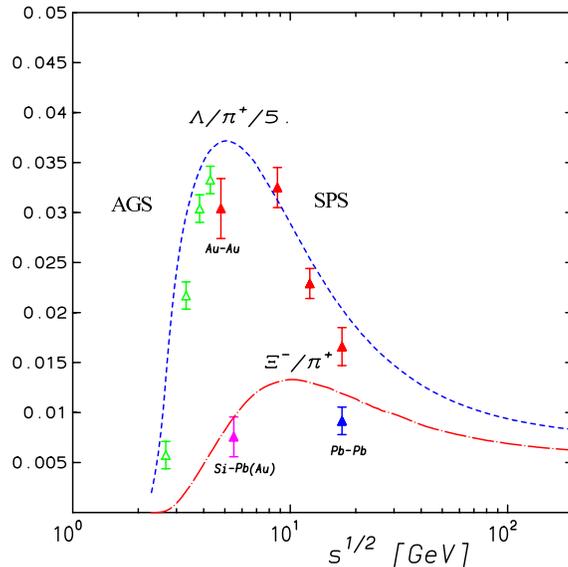}
\vspace*{-0.5cm}
\caption[]{Energy dependence of Lambda and cascade multiplicity relative
to positive pion multiplicity, at AGS and CERN SPS energy, Ref.1,
with statistical model interpolation, Ref.32.}
  \end{center}
%\label{fig1}
\end{figure}

Recent work with the grand canonical hadronization model, by
Braun-Munzinger, Cleymans, Oeschler, Redlich and Stachel \cite{32}
has shown that such, at first sight remarkable, nonmonotoneous
behaviour is, again, semiquantitatively captured. They interpolate
among the various GCE fits at increasing $\sqrt{s}$, to obtain a
continuous hadronic freezeout trajectory in the $T$, $\mu$ plane.
The result is shown in Fig. 8 to reproduce the overall features of
the data. Furthermore they showed that these separate strange to
nonstrange $\Lambda$ or $K^+$ to pion yield dependences on
$\sqrt{s}$ are the consequence of a more general maximum in the
"Wroblewski-ratio", $\lambda \equiv 2(s+ \overline{s})/(u +
\overline{u}+d+ \overline{d})$ at similar $\sqrt{s}$. This finding
is illustrated in Fig. 9. The Wroblewski $\lambda$ dependence on
$T$ and $\mu_B$ is shown in the plane of $T, \: \mu_B$ in a set of
curves each corresponding to a fixed value of $\lambda$, from 0.3
to 0.8, as obtained from the GCE model. The (dashed) hadronic
freeze-out curve intersects these lines, steeply at first, from
high $\mu_B$ downwards ("strangeness enhancement") to $\mu_B
\approx$ 450 MeV where $\sqrt{s}$=6-8 GeV is implied. It peaks
there at $\lambda$ =0.65 indicating a maximum global strangeness
to nonstrangeness ratio, the reflection of which we saw in Fig. 8.
Then $\lambda$ falls back to 0.4 toward top SPS and RHIC energies.

\begin{figure}[ht]
\begin{center}
%\vspace*{-1.1cm}
                 \insertplot{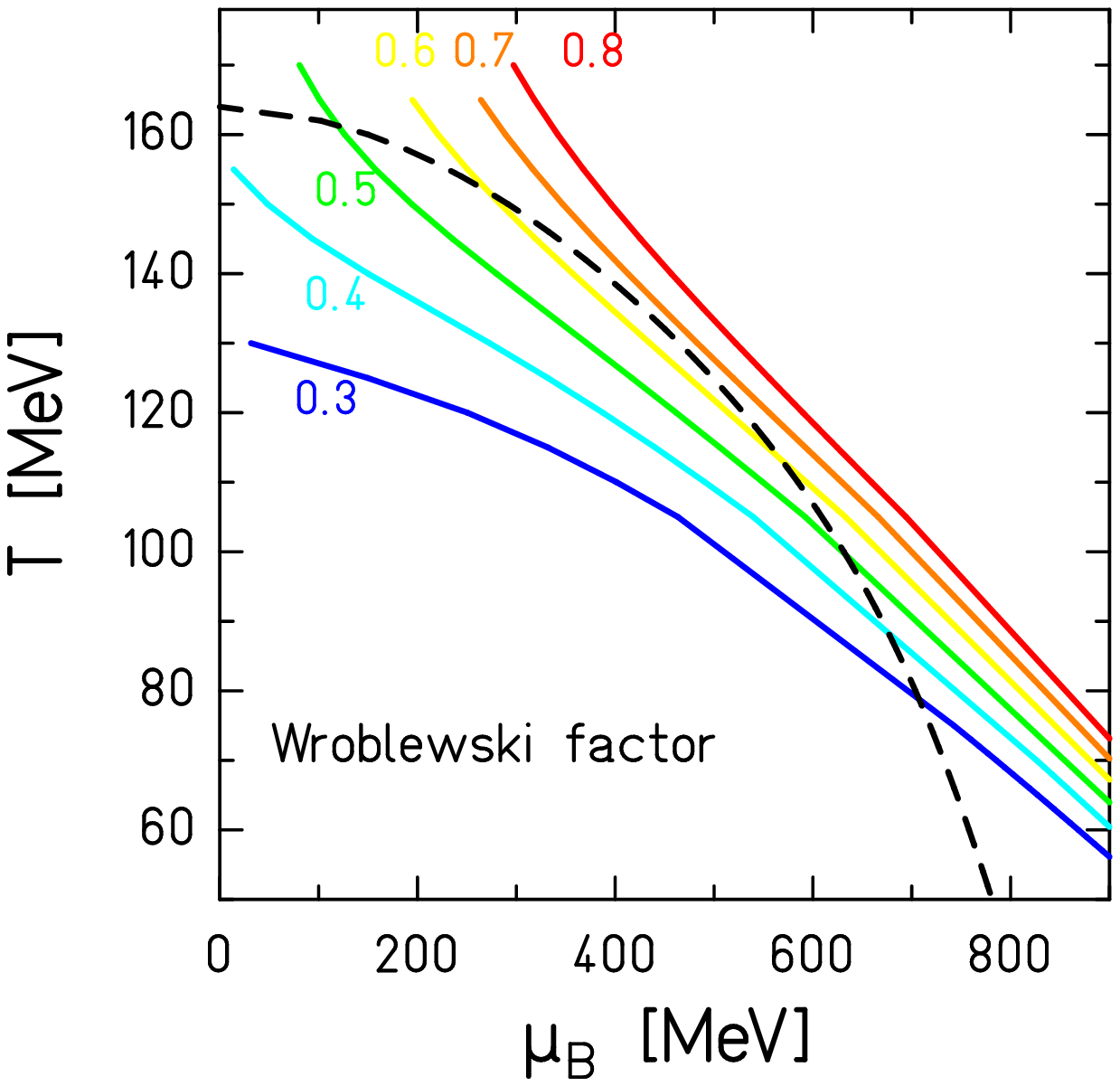}
%\vspace*{-2.1cm}
\caption[]{Lines of constant Wroblewski $\lambda$ parameter in the $T, \: \mu_B$
plane in the GCE model, intersected by the hydrostatic chemical freezeout curve
(dashed), from Ref.32.}
%\label{fig1}
\end{center}
\end{figure}

Cleymans has shown \cite{33} that this general evolution of the
strange to nonstrange hadronic population is the consequence, at
the level of the statistical model, of qualitatively different
trends concerning the basic parameters. While $\mu_B$ drops to
zero continuously with increasing $\sqrt{s}$ the apparent
hadronization temperature  turns into saturation (after a similar
steeply ascending passage initially) toward $T=170 \pm$10 MeV,
above about $\sqrt{s}$= 6-8 GeV where it has already reached
$T=140-150$ MeV. This picture can actually be even recast in the
terminology of microscopic collisions: at lower $\sqrt{s}$
strangeness production is "encouraged" by associated production
channels owing to the prevailing high net baryon density, then to
turn over to $s \overline{s}$ production from free fireball energy
\cite{34}. This picture would, alone by itself, indicate merely a
saturation of the relative strangeness yield.  An additional
feature sets in with the advent of limiting hadron temperature
which can not be understood from a continuous evolution of
hadronic collision energy alone. It signifies the advent of
partonic phase dynamics. Actually, "no reasonable person would
doubt that toward $\sqrt{s}$ = 200 GeV the most simple picture
arises from the interaction of quarks and gluons", to quote Lerry
McLerran \cite{35}.

Thus, in concluding this report, let me recapitulate that the
Hagedorn statistical model does not {\it predict} anything except
for the existence of a limiting temperature. It {\it reacts} to
the data in its particular language, and from this reflection we
infer that hadron multiplicities obey a grand
canonical order which indicates that an extended, collectively
interacting "fireball" of strongly interacting matter is formed in
AA collisions, that merges closely with the QCD phase
transformation boundary toward top SPS, and RHIC energy, thus
providing for an estimate of the QCD critical temperature. At
lower $\sqrt{s}$ the overall behaviour of the hadronic multiplicities
may reflect the advent of this phase boundary, but at present we
lack an appropriate understanding of dense hadronic matter
dynamics to fully comprehend the apparent validity of the grand
canonical model, also in this domain.

\end{document}